\documentclass{aa}
\usepackage{epsfig}
\usepackage{graphics}

\newcommand{\be}{\begin{equation}}
\newcommand{\ee}{\end{equation}}

\begin{document}

\title{ Dark energy and the structure of the Coma
cluster of galaxies }

\author{A. D. Chernin\inst{1,2}   \and  G.S.~Bisnovatyi-Kogan\inst{3}
\and P. Teerikorpi\inst{1} \and M. J.
Valtonen\inst{1}  \and G. G. Byrd\inst{4} \and M. Merafina\inst{5} }

\institute{Tuorla Observatory, Department of Physics and Astronomy, University of Turku, 21500 Piikki\"{o},
Finland \and Sternberg Astronomical Institute, Moscow
University, Moscow, 119899, Russia  \and Space Research Institute, Russian Academy of Sciences, Moscow,
Russia
\and University of Alabama,
Tuscaloosa, AL 35487-0324, USA  \and Department of Physics, University of Rome "La Sapienza", Rome, Italy }

\authorrunning{A.D. Chernin et al.}
\titlerunning{Dark energy and the Coma cluster}

\date{Received / Accepted}

\abstract
{We consider
the Coma cluster of galaxies as a gravitationally bound physical
system embedded in the perfectly uniform static dark energy
background as implied by the $\Lambda$CDM cosmology.
}
{ We ask if the density of  dark energy is high enough to affect the
structure of a large rich cluster of galaxies?
}
{We use recent
observational data on the cluster together with our theory
of local dynamical effects of dark energy, including the zero-gravity radius
$R_{\rm ZG}$ of the local force field as the key parameter.
}
{1) Three  masses are defined which characterize the structure of a regular cluster:
the matter mass $M_{\rm M}$, the dark-energy effective mass $M_{\rm DE}$ ($<0$) and the gravitating mass $M_{\rm G}$
($= M_{\rm M} + M_{\rm DE}$). 2) A new
matter density profile is suggested which reproduces well the
observational data for the Coma cluster in the radius range
from 1.4 Mpc to 14 Mpc and takes into account the dark energy background. 3)
Using this profile, we calculate upper limits for the total size of the Coma cluster, $R \le R_{\rm ZG} \approx
20$ Mpc, and its total matter mass, $M_{\rm M} \la M_{\rm M}(R_{\rm ZG}) = 6.2 \times
10^{15} M_{\odot}$.
}
{ The dark energy antigravity affects strongly the
structure of the Coma cluster at large radii $R \ga 14$
Mpc and should be taken into account when its total mass is derived. }


\maketitle

\section{Introduction}

The Coma Cluster of galaxies (A1656) is the most massive well-studied
regular gravitationally bound aggregation of matter  in the observable Universe. In his classic work Zwicky (1933, 1937)
 applied the virial theorem to the
cluster and showed that dark matter dominated the
system.
Zwicky estimated its mass as $3
\times 10^{14} M_{\odot}$, when normalized to the
Hubble constant $h = 0.71$ used hereafter

Decades
later, The \& White (1986) found an order of magnitude larger
value, $2 \times 10^{15} M_{\odot}$, with  a generalization of
the virial theorem which must be used when the observed sample does not include
the entire system. Hughes (1989, 1998) obtained a similar value
$(1-2) \times 10^{15} M_{\odot}$ with X-ray data under the
assumption that the cluster hot intergalactic gas is in
hydrostatic equilibrium. In a similar way, Colles (2006)
reports the mass $4.4 \times 10^{14} M_{\odot}$ inside the radius
of 1.4 Mpc. A weak-lensing analysis gave $2.6 \times
10^{15} M_{\odot}$ (Kubo et al. 2007) within 4.8 Mpc radius.
Geller et al. (1999, 2011) extended mass estimates to the
outskirts of the cluster using the caustic technique (Diaferio \&
Geller 1997, Diaferio 1999) and found the mass $2.4 \times 10^{15}
M_{\odot}$ within the 14 Mpc radius. Here the $2\sigma$
error is $1.2 \times 10^{15} M_{\odot}$, hence it does not contradict the apparently larger
mass within 4.8 Mpc.

In this paper, we re-examine the matter mass estimation of the Coma
cluster using the data above and a new theory model
which describes the cluster as a bound spherical system embedded
in the cosmic background of dark energy. We find that dark energy
affects strongly the cluster structure at large distances $R \ge
14$ Mpc from the cluster center and must be taken into account
in the matter mass estimate.

Basic theory is
outlined in Sect.2, three characteristic masses of a  regular
cluster are introduced  in Sect.3, a new matter mass
profile is defined in Sect.4, the upper bound on the total size
of the Coma cluster is calculated and discussed in Sect.5, and the results are summarized
in Sect.6.

\section{Local dynamical effects of dark energy}

We adopt  the $\Lambda$CDM cosmology which identifies
dark energy with Einstein's cosmological constant $\Lambda$ and
treats it as a perfectly uniform vacuum-like fluid with
the constant density $\rho_{\rm DE} = 0.71 \times 10^{-29}$ g
cm$^{-3}$.
The
dark energy background produces antigravity which is stronger than
the matter gravity in the present Universe as a whole. This makes the
cosmological expansion accelerate as discovered
by Riess et al. (1998) and  Perlmutter et al. (1999).

The cosmic
antigravity can be stronger than gravity not only  globally,
but also locally on the scales of $\sim 1-10$ Mpc (Chernin et al. 2000, 2006; Chernin 2001; Byrd et al. 2007, 2012),
as studied using the HST observations made by
Karachentsev's team (e.g., Chernin et al. 2010, 2012).

The local weak-field dynamical effects of dark energy can be
adequately described in terms of Newtonian mechanics (e.g., Chernin 2008).
Such an approach borrows from General Relativity the major result:
the effective gravitating density of a uniform medium is given by
the sum
\be \rho_{\rm eff} = \rho + 3 P, \ee
\noindent where $\rho$ and $P$ are the fluid's density and pressure ($c = 1$ hereafter). In the
$\Lambda$CDM model, the dark energy equation of state is $P_{\rm
DE} = - \rho_{\rm DE}$, and its effective gravitating density
\be \rho_{\rm DEeff} = \rho_{\rm DE} + 3 P_{\rm DE} = - 2 \rho_{\rm DE} < 0\ee
\noindent is negative, producing antigravity.
Einstein's "law of universal antigravity" says that a point mass $M$
within uniform dark energy  generates an acceleration $a(r)$ which includes
in addition to the Newtonian term $a_{\rm N}(r) = -GM/r^2$ the antigravity effect
of dark energy
\be a_{\rm E}(r) = - {\frac{4\pi G}{3}}\rho_{\rm DEeff}r^3/r^2 = +
{\frac{8\pi G}{3}}\rho_{\rm DE}r. \label{Eq.3} \ee

\noindent
Then a test particle at the distance $R$
from the center of a spherical matter mass $M_{\rm M}$ (beyond the mass) has the radial acceleration
in the reference frame of the mass center:
\be a (R) = a_{\rm N} (R) + a_{\rm E} (R) = - G \frac{M_{\rm M}}{R^2} +
{\frac{8\pi G}{3}}\rho_{\rm DE}R. \label{Eq.4} \ee

\noindent Eq.4 comes  from the Schwarzschild-de Sitter
spacetime in the weak field approximation.
The net {\bf acceleration} $a(R)$ is zero at the distance (Chernin et
al. 2000, Chernin 2001, 2008)
\be R  =  R_{\rm ZG} = [\frac{M_{\rm M}}{{\frac{8\pi}{3}}\rho_{\rm
DE}}]^{1/3} = 11(\frac{M_{\rm M}}{10^{15} M_{\odot}})^{1/3}\, {\rm Mpc}\,,
\label{Eq.5} \ee

\noindent Gravity dominates at distances $R <
R_{\rm ZG}$, while antigravity is stronger than gravity at $R
> R_{\rm ZG}$. A gravitationally bound system
with the mass $M_M$ can exist only inside its zero-gravity
sphere of the radius $R_{\rm ZG}$.

\section{Three masses of a regular cluster}

The presence of dark energy in the spherical volume of a regular
cluster like Coma  may be quantified by its effective
gravitating mass within a given clustrocentric
radius $R$:

 \be M_{\rm DE} (R)  =
-\frac{8\pi}{3} \rho_{\rm DE} R^3
= - 0.85 \times 10^{15} [\frac{R}{10
{\rm Mpc}}]^3 M_{\odot}
\ee

\noindent The matter (dark matter and baryons) content of the cluster is
characterized by the mass $M_{\rm M} (R)$
inside  radius $R$:
\be M_{\rm M} (R) = 4 \pi \int \rho(R) R^2 dR. \ee

\noindent Here $\rho (R)$ is the matter density within the
cluster. The sum
\be M_{\rm G} (R) = M_{\rm M} (R) + M_{\rm DE}(R), \ee

\noindent is the total gravitating mass within the radius $R$.
It is this mass that can be directly
measured by the methods cited in
Sec.1, which are all related to gravitation and give the gravitating
mass $M_{\rm G}(R)$, rather than the matter mass $M_{\rm M}(R)$ which,
however, can be derived from the data of Sec.1 using
Eq.8: $M_{\rm M} = M_{\rm G} - M_{\rm DE}$. So one has for $R
= 1.4$ Mpc:
\be M_{\rm DE} = - 2.3 \times 10^{12} M_{\odot},  \;\;\;     M_{\rm M}
\simeq M_{\rm G} = 4.4 \times 10^{14} M_{\odot}.\ee

For $R = 4.8$ Mpc:
\be M_{\rm DE} = - 9.4 \times 10^{13} M_{\odot}, \;\;\;     M_{\rm M} \simeq
M_{\rm G} = 2.6 \times 10^{15} M_{\odot}.\ee

And for $R = 14$ Mpc:
\be M_{\rm DE} = - 2.3 \times 10^{15} M_{\odot}, \;\;\;    M_{\rm M} = 4.7
\times 10^{15} M_{\odot}.\ee

As we see, in the inner cluster at $R = 1.4$
and 4.8 Mpc the dark energy contributes practically
nothing compared to the gravitating mass, so $M_{\rm G} \simeq
M_{\rm M}$ here. But (curiously) the absolute
value of the dark energy mass $M_{\rm DE}$  nearly equals
the gravitating mass $M_{\rm G}$ at $R = 14$ Mpc; as a result, the
matter mass $M_{\rm M} \simeq 2 M_G \simeq 2|M_{\rm DE}|$ at this radius.
The difference between the Eq.11 estimate and the observed
value of $M_{\rm G}$ is at the level of $4\sigma$ ($1 \sigma \approx
\frac{1}{4} M_{\rm G}$, Sec.1).

In the outer regions of the Come cluster,
\be |M_{\rm DE}| > M_{\rm G},  \;\;\; R > 14 \;{\rm  Mpc},\ee

\noindent where the antigravity effect thus is significant.

\section{Matter mass profile}

Our estimate of the Coma matter mass within $R=14$
Mpc (Eq.11) may be compared with estimates following from
traditional matter density profiles for dark haloes.

\subsection{NFW and Hernquist profiles}

 The popular NFW profile (Navarro et al. 2005) is
\be \rho = \frac{4\rho_s}{\frac{R}{R_s}(1 + \frac{R}{R_s})^2},
\label{Eq.13}
\ee

\noindent where $R$ is again the distance from the cluster center,
$\rho_s = \rho(R_s)$ and $R_s$ are constant parameters. At small
radii, $R << R_s$, the matter density goes to infinity, $\rho
\propto 1/R$ as $R$ goes to zero; at large distances, $R >> R_s$,
the density slope is $\rho \propto 1/R^3$. With this profile, the
matter mass profile is
\be M_{\rm M} (R) = 16 \pi \rho_s R_s^3 [\ln (1 + R/R_s) -
\frac{R/R_s}{1 + R/R_s}],  \label{Eq.14} \ee

To find the parameters $\rho_s$ and $R_s$, we use the
small-radii data of Sec.1: $M_1 = 4.4 \times 10^{14} M_{\odot}$ at
$R_1 = 1.4$ Mpc, $M_2 = 2.6 \times 10^{15} M_{\odot}$ at $R_2 =
4.8$ Mpc. At these radii, the gravitating masses are nearly
equal to the matter masses there (c.f. Sect.3).
The values of $M_1, R_1$ and $M_2, R_2$ together with
Eq.14 lead to two logarithmic equations for the
two parameters of the profile, which can easily be solved:
$ R_s = 4.7 \; \; {\rm Mpc}, \;\;\; \rho_s = 1.8 \times 10^{-28} \;\;
{\rm g/cm}^3$.
Then we find the matter mass within $R = 14$ Mpc:
\be M_{\rm M} \simeq 8.7 \times 10^{15}\; M_{\odot}\,, \ee

\noindent considerably larger (over 70\%) than  given by Eq.11.

Another widely-used density profile (Hernquist 1990) is
\be \rho(R) \propto \frac{1}{R (R + \alpha)^3}.
\label{Eq.17}
\ee

\noindent Its small-radius behavior is the same as in the NFW
profile: $\rho \rightarrow \infty$, as $R$ goes to zero. The
slope at large radii is different: $\rho \propto 1/R^4$. The
corresponding mass profile is
\be M_{\rm M} (R) = M_0 [\frac{R}{R + \alpha}]^2.
\label{Eq.18}
\ee
\noindent The  parameters $M_0$ and $\alpha$ can be found
from the same data as above on $M_1, R_1$ and $M_2,
R_2$:
$ M_0 = 1.4 \times 10^{16} \; M_{\odot}, \;\; \alpha = 6.4 \;
{\rm Mpc}$, giving another value for the mass within $14$ Mpc:
\be M_{\rm M} = 6.6 \times 10^{15} M_{\odot}, \;\; R =  14 \; {\rm Mpc},
\ee

\noindent Now the difference from the figure of Eq.11 is about 40\%.

\begin{figure}
\epsfig{file=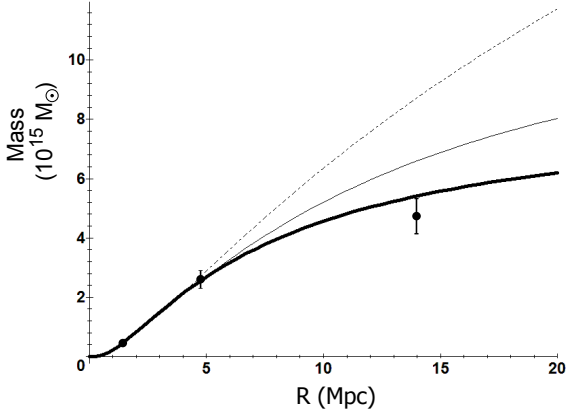, angle=0, width=8.0cm} \caption{Matter mass $M$
vs. radius $R$ for three profiles. 1) NFW mass profile of Eq.14:
$M =  13.8 [\ln(1+R/4.7)-\frac{R}{4.7+R}]$ - dashed line; 2)
Hernquist profile of Eq.17: $M = 14\left(\frac{R}{6.4+R}\right)^2$
- thin line; 3) New mass profile of Eq.19: $M = 8.7
\left(\frac{R}{2.4+R}\right)^3$ - thick line. Here $R$ is in Mpc
and $M$ is in $10^{15}M_{\odot}$. The values of the matter mass at
radii $R = 1.4, 4.8, 14$ Mpc are also showed as given by Eqs.9-11
with $1\sigma$ error bars. } \label{fig1}
\end{figure}

\subsection{A modified mass profile}

In a search for a more suitable mass profile for the Coma cluster,
we  try a modified version of Hernquist's relation:
\be M_{\rm M} (R) = M_* [\frac{R}{R + R_*}]^3.
\label{Eq.21}
\ee

\noindent The power of 3 is used now instead of the power of 2 in
Eq.18. This mass profile comes from the density profile:
\be \rho = \frac{3}{4\pi} M_* R_* (R + R_*)^{-4}.
\label{Eq.22}
\ee

\noindent The density goes to a constant as $R$ goes to zero; at
large radii $\rho \propto 1/R^4$, like in Hernquist's profile.

The parameters $M_*$ and $R_*$ are found again from the data for
the radii of 1.4 and 4.8 Mpc:
$ M_* = 8.7 \times 10^{15} \; M_{\odot}, \;\; R_* = 2.4 \;
{\rm Mpc}$.
The new profile leads to a smaller matter mass at $14$ Mpc:
\be M_{\rm M} = 5.4 \times 10^{15} M_{\odot},\ee

\noindent which is equal to the  Eq.11 value within 15\%
accuracy.
 The three profiles are graphed in Figs.1 and 2.

\begin{figure}
\epsfig{file=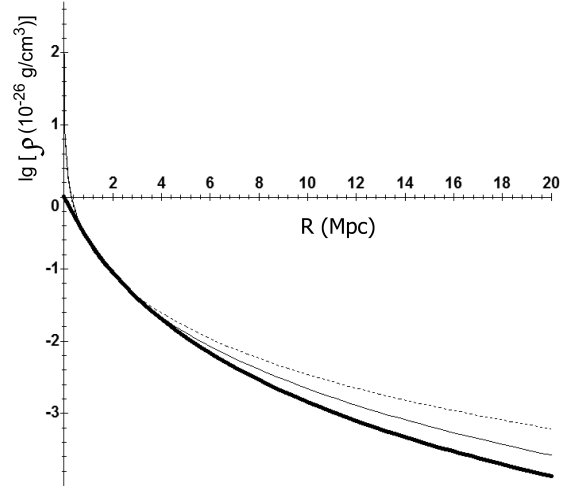, angle=0, width=8.0cm}
\caption{Matter density  $\rho$ vs.
radius $R$. 1) NFW density
profile (Eq.13): $\rho = \frac{7.5}{R(4.7+R)^2}$ - dashed line; 2) Hernquist profile (Eq.16): $\rho=
\frac{97}{R(6.4+R)^3}$  - thin
line; 3) New density profile (Eq.20): $\rho =
\frac{34}{(2.4+R)^4}$ - thick line.
}
\label{fig2}
\end{figure}

\section{Discussion}

We now discuss some implications of the above results.

\subsection{Upper limits to size and mass}

The strong effect of dark energy at large radii
puts an absolute upper limit on the total size of the cluster:
the system can be gravitationally bound only if gravity
dominates in its volume (as we mentioned in Sec.2). In
terms of the three different masses, this criterion may
be given in the form
\be M_{\rm G} \ge 0, \;\;\; M_{\rm M} \ge |M_{\rm DE}|. \ee

\noindent Both inequalities are met, if the
system is not larger than its zero-gravity radius ( Eq.5):
$R \le R_{\rm max} = R_{\rm ZG}$.

\noindent If the radius of a  system with matter mass $M_{\rm M}$
is equal to the maximal radius $R = R_{\rm max}$, its mean matter density
(see Bisnovatyi-Kogan \& Chernin 2012) is
\be \langle \rho_{\rm M} \rangle = \frac{M_{\rm M}}{\frac{4\pi}{3} R_{\rm ZG}^3} =
2 \rho_{\rm DE}. \label{eq26} \ee

\noindent This relation and the new profile (Eq.19) now lead to  $R_{\rm max}$ and
the corresponding matter mass, $M_{\rm max}=M_{\rm M}(R_{\rm max})$:
\be R_{\rm max} = R_{\rm ZG} = 20\; {\rm Mpc}, \;  M_{\rm max} =
M_{\rm M} (R_{\rm ZG}) = 6.2 \times 10^{15} M_{\odot}.
 \label{eq27} \ee

For comparison, the other  profiles lead to
\be R_{\rm max} = 25\; {\rm Mpc}, \;\; M_{\rm max} = 1.5 \times
10^{16} M_{\odot}. \; ({\rm NFW})
 \label{eq28} \ee
\be R_{\rm max} = 22\; {\rm Mpc}, \;\; M_{\rm max} = 9.1 \times
10^{15} M_{\odot}. ({\rm Hernquist})
 \label{eq29} \ee

\subsection{Close to the maximal size?}

Our studies of nearby systems like the
Local Group and the Virgo and Fornax
clusters (e.g., Chernin et al. 2010, 2012a) suggest their sizes
are not far from the
zero-gravity radii. Around them, flows of galaxies
are seen (Karachentsev et al.  2009;
2010, Nasonova et al. 2011), and
the systems are located in
the gravity-dominated regions ($R < R_{\rm ZG}$) and the outflows are
at $R > R_{\rm ZG}$.
If the local systems have nearly maximal sizes, this may
explain the apparent underdensity of the local universe (Chernin et al. 2012b).

It is tempting to ask if the matter distribution
could extend to somewhere near the maximal distance of 20 Mpc in the Coma cluster as
well. If so, its mass would be near the
upper limit evaluated above, and still consistent with
the theory of
large-scale structure formation which claims the range $2 \times
10^{15} < M < 10^{16} M_{\odot}$ for the most massive bound
objects in the Universe (Holz \& Perlmutter 2010, Busha et al.
2005).
 Another implication is the
predicted (eq. 23) mean matter density of the
system $=$ twice the dark energy density, which does not depend
on the density profile assumed (Merafina et al. 2012;
Bisnovatyi-Kogan \& Chernin 2012). Its observational confirmation
would directly indicate the key role of dark energy
for the structure of the system. Using the cosmological
matter density parameter $\Omega_{\rm m} =  0.27$, its mean density contrast would be
\be \delta = \frac{\langle \rho \rangle - \rho_{\rm m}}{\rho_{\rm m}} =
\frac{2 \Omega_{\rm DE}}{\Omega_{\rm m}} - 1 = 4.2\,.
 \label{eq30} \ee

Another general prediction (Chernin et al. 2006) is that at distances $R > R_{\rm ZG}$
any galaxy in the outflow should have a velocity higher than
\be V_{\rm esc} = H_{\Lambda} R [1 + 2 (R_{\rm ZG}/R)^3 - 3
(R_{\rm ZG}/R)^2]^{1/2}
 \label{eq33}
\ee

\noindent Here
$ H_{\Lambda} =  [\frac{8 \pi G}{3}\rho_{\rm DE}]^{1/2} = 61\;
{\rm km s}^{-1} {\rm Mpc}^{-1}$
 depends on the dark energy density only.
Furthermore, within the simplified model of Einstein-Straus vacuole,
one expects the flow to reach the global Hubble rate at
the edge of the vacuole ($\approx 1.7R_{\rm ZG}$, Teerikorpi \& Chernin 2010;  Hartwick 2011).

The situation is complicated by
the fact that the Coma cluster is not isolated, but lies within the CfA Great Wall.

\subsection{Dark energy estimator}

We have assumed that the dark energy density
inside the Coma cluster is equal to its global value.
One can reverse the argument and
consider the local dark energy density as an unknown constant
$\rho_x$. Its value may be independently estimated using
our concept of three cluster masses.

The data (Geller et al. 1999, 2011) give the gravitating
mass $M_{\rm G}$ within $R = 14$ Mpc. The mass $M_{\rm M}$ at
the same radius may be found by extrapolating the data
from $R = 1.4$ Mpc and $R = 4.8$ Mpc using a reasonable
matter density profile (Sect.4). Then with the masses $M_{\rm G}$
and $M_{\rm M}$ known for $R = 14$ Mpc, we may find the dark energy
mass, $M_{\rm DE}(R) = M_{\rm G} - M_{\rm M}$, at the same radius. Finally, the
dark energy density inside the cluster is estimated from
\be \rho_x = |M_{\rm G} - M_{\rm M}| \frac{1}{\frac{8\pi}{3} R^3}; \;\; R = 14
\; {\rm Mpc}. \ee

With the density profiles of  Eqs. 13 and 16 in Sect.4, Eq.29 gives for the local
dark energy density the values
\be \rho_x = (1.2-2) \times 10^{-29} \;{\rm  g \; cm}^{-3}, \ee

\noindent  equal to the global value $\rho_{\rm DE}$ within an order of magnitude
accuracy (to avoid circular argumentation, we do not use the profile of Eq.20 here, as it was
partly suggested from considerations related to the global value itself).

\section{Conclusions}

1. Three masses which characterize the structure of a regular cluster
(like Coma) are defined as functions of the
radius $R$: the matter (dark matter and
baryons) mass $M_{\rm M} (R)$, the dark-energy effective gravitating
mass $M_{\rm DE}$ (negative), and the total gravitating mass
$M_{\rm G} (R) = M_{\rm M} + M_{\rm DE}$.

2. Of these masses, only the gravitating mass $M_{\rm G}$ reveals
itself directly in observations at various distances from the
cluster center. The dark energy mass $M_{\rm DE}$ may
be derived using the known global value of the dark energy density.

3. The mass $2.4 \times 10^{15} M_{\odot}$ measured at $R = 14$
Mpc by Geller et al. (1999, 2011) is the
gravitating mass $M_{\rm G}$ inside this radius. The corresponding matter mass is $M_{\rm M}
\simeq 2 M_{\rm G}$.

4. At small radii, $R << 14$ Mpc, dark energy effects are almost
negligible, $|M_{\rm DE}| << M_{\rm M}$, and the gravitating mass $M_{\rm G}$ is
practically equal to the matter mass $M_{\rm M}$. At large radii $R
\ge 14$ Mpc, the antigravity effects are strong and $|M_{\rm DE}| \ge
M_{\rm G}$.

5. A new matter mass profile for the Coma cluster
reproduces well the observational data and accounts for the
dark energy effects in the radius range from 1.4 to 14 Mpc and
beyond: $M_{\rm M} (R) = M_* [\frac{R}{R + R_*}]^3$, where the constants $M_*$ and
$R_*$ can be found from the data for
small radii.

6. The available observational data and the new mass profile give
upper limits for the Coma cluster total size, $R \la 20$ Mpc,
and total matter mass, $M_{\rm M} \la 6.2 \times 10^{15} M_{\odot}$.

\acknowledgements We thank Yu.N. Efremov, I.D. Karachentsev, D.I.
Makarov, O.G. Nasonova and A.V. Zasov for many useful discussions.
A.C. appreciates a partial support from the RFBR grant
13-02-00137. The work of G.B.-K. was partly supported by RFBR
grant 11-02-00602, the RAN Program "Formation and evolution of
stars and galaxies", and Russian Federation President Grant for
Support of Leading Scientific Schools NSh-3458.2010.2.


\begin{thebibliography}{}







\item Bisnovatyi-Kogan G.S. \& Chernin, A.D. , 2012, Ap\&SS 338, 337\\

\item Busha M.T., Evrard, A.E., Adams, F.C., Wechsler, R.H., 2005, MNRAS, 363, L11\\


\item Byrd G.G., Chernin A.D. and Valtonen M.J., 2007, {\it Cosmology:
Foundations and Frontiers} Moscow, URSS\\

\item Byrd G.G., Chernin A.D., Teerikorpi P., Valtonen M.J., 2012, {\it
Paths to Dark Energy: Theory and Observation} N.Y., de Gruyter\\

\item Chernin A.D., 2001, Physics-Uspekhi 44, 1099\\

\item Chernin A.D., 2008, Physics-Uspekhi 51, 267\\

\item Chernin A.D., Teerikorpi P., Baryshev Yu.V., 2000,
[astro-ph/0012021] Adv. Space Res., 31, 459, 2003\\

\item Chernin A.D., Teerikorpi P., Baryshev Yu.V., 2006, A\&A 456, 13\\





\item Chernin A.D., Karachentsev I.D., Nasonova O.G., et al., 2010, A\&A
520, 104\\

\item Chernin A.D., Teerikorpi P., Valtonen M.J. et al. 2012a, Astr. Rep. 56, 653\\

\item Chernin A.D., Teerikorpi, P., Valtonen, M.J. et al., 2012b, A\&A 539, A4\\

\item Colles M., 2006, in {\it Encyclopedia of Astronomy and
Astrophysics}, P.Murdin,
ed., IOP Pubs.Ltd, UK, p.245\\


\item Diaferio, A., Geller M.J., 1997, ApJ, 481, 633\\

\item Diaferio A., 1999, MNRAS, 309, 610\\






\item Geller M.J., Diaferio A., Kurtz, M.J.,1999, ApJ, 517, L23\\

\item Geller M.J., Diaferio A., Kurtz, M.J., 2011, AJ 142, 143\\



\item Hartwick F.D.A., 2011, AJ 141, 198\\

\item Hernquist L., 1990, ApJ, 356,  359\\




\item Holz D.E. \& Perlmutter S., 2010, arXiv:1004.5349\\

\item Hughes, John P., 1989, ApJ 337, 21   \\

\item Hughes, J. P., 1998, "Untangling Coma Berenices: A New Vision of an
Old Cluster", Proceedings of the meeting held in Marseilles,
June 17-20, 1997, Eds.: Mazure, A., Casoli F., Durret F. , Gerbal D.,
Word Scientific, p 175.\\











\item Karachentsev I.D., Kashibadze O.G., Makarov D.I., et al., 2009, MNRAS 393, 1265\\


\item Kubo J.M., Stebbins A., Annis, J. et al., 2007, ApJ, 671, 1466\\



\item Merafina M., Bisnovatyi-Kogan G.S., Tarasov S.O., 2012, A\&A, 541, 84\\

\item Nasonova O.G., de Freitas Pacheco, J.A., Karachentsev I.D., 2011, A\&A, 532, 104\\

\item Navarro J., Frenk C.S., White S.D.M., 2005, ApJ 463, 563\\


\item Perlmuter S., Aldering G., Goldhaber G. et al., 1999, ApJ 517, 565\\


\item Riess A.G., Filippenko A.V., Challis P., et al., 1998, AJ 116, 1009\\


\item Rines K., Geller, M.J., Diaferio, A., Kurtz, M.J., Jarrett, T.H., 2004, AJ, 128,  1078\\

\item Rines K. \& Diaferio A., 2006, AJ 132, 1275\\







\item Teerikorpi P., Chernin A.D., 2010, A\&A 516, 93\\

\item The L. S., White S.D.M., 1986, AJ, 92, 1248\\









\item Zwicky, F., 1933, Helvetica Phys. Acta, 6, 110\\

\item Zwicky F., 1937,  ApJ, 86, 217

\end{thebibliography}
\end{document}